\documentclass[a4paper,12pt]{article}

\usepackage{jheppub}

\title{\boldmath A note on background independence in $\text{AdS}_3$ string theory}

\abstract{In this note, we comment on the path integral formulation of string theory on $\mathcal{M}\times\text{S}^3\times\mathbb{T}^4$ where $\mathcal{M}$ is any hyperbolic 3-manifold. In the special case of $k=1$ NS-NS flux, we provide a covariant description of the worldsheet theory and argue that the path integral depends only on the details of the conformal boundary $\partial\mathcal{M}$, making the background independence of this theory manifest. We provide a simple path integral argument that the path integral localizes onto holomorphic covering maps from the worldsheet to the boundary. For closed manifolds $\mathcal{M}$, the gravitational path integral is argued to be trivial. This implies that the bulk gravitational theory has precisely one state in its Hilbert space. Finally, we comment on the effect of continuous deformations of the worldsheet theory which introduce non-minimal string tension.}

\author{Bob Knighton}
\affiliation{Department of Applied Mathematics \& Theoretical Physics, University of Cambridge,\\
Wilberforce Road, Cambridge CB3 0WA, United Kingdom}
\emailAdd{rik23@cam.ac.uk}

\usepackage{bm} 
\usepackage[dvipsnames]{xcolor}
\definecolor{green_maf}{RGB}{28, 166, 46}
\definecolor{blue_mrg}{RGB}{12, 143, 145}
\definecolor{detail}{RGB}{110,110,110}
\usepackage{amsmath} 
\usepackage{nccmath} 
\usepackage{amssymb} 
\usepackage{amsthm} 
\usepackage{mathtools} 
\usepackage[utf8]{inputenc} 
\usepackage{braket}
\usepackage{enumerate}
\usepackage[many]{tcolorbox}
\usepackage[inline]{enumitem}
\usepackage{comment}
\usepackage{soul} 
\usepackage{tikz}
\usepackage{csquotes}
\usepackage{mathrsfs}
\usepackage{tensor}

\newtcolorbox{empheqboxed}{colback=gray!30, 
 colframe=white,
 width=\textwidth,
 sharpish corners,
 top=-2mm, 
 bottom=0pt
}

\hypersetup{
		pdfencoding=unicode,
		colorlinks=true,
		urlcolor=Maroon,
		linkcolor=RoyalBlue,
		citecolor=Maroon,
		pdfstartview=FitH,
		linktocpage=true
}

\usetikzlibrary{calc,decorations.markings}
\usetikzlibrary{decorations.pathmorphing}
\usetikzlibrary{shapes,backgrounds}
\usetikzlibrary{fadings}
\usetikzlibrary{cd}
\usetikzlibrary{decorations.pathreplacing,calligraphy}
\usetikzlibrary{hobby}

\tikzset{
	partial ellipse/.style args={#1:#2:#3}{
		insert path={+ (#1:#3) arc (#1:#2:#3)}
	}
}

\tikzset{
  every overlay node/.style={
    draw=black,fill=white,rounded corners,anchor=north west,
  },
}

\tikzfading
[
  name=fade out,
  inner color=transparent!0,
  outer color=transparent!100
]

\newif\ifdetails
\detailstrue

\notoc

\begin{document}

\maketitle

\newpage

\section{Introduction}

For the last three decades, the AdS/CFT correspondence \cite{Maldacena:1997re,Witten:1998qj} has served as a guiding principle for the study of quantum gravity. The most basic statement of the holographic dictionary is that the quantum gravity path integral on an asymptotically $\text{AdS}_{d+1}$ manifold $\mathcal{M}$ should be equal to the path integral of a corresponding $\text{CFT}_d$ on the conformal boundary $X$. Mathematically\footnote{In the quantum gravity literature it is common to express the left-hand-side as a sum over \textit{all} suitable geometries with boundary $X$. In such cases, we consider the sum over geometries as a part of the \textit{definition} of $Z_{\text{grav}}$ and therefore do not write the sum out explicitly.}
\begin{equation}\label{eq:holographic-dictionary}
Z_{\text{grav}}(\mathcal{M})=Z_{\text{CFT}}(X)\,.
\end{equation}
Perhaps the most striking, and poorly understood, feature of this equation is the independence of the right-hand-side on the details of the bulk manifold $\mathcal{M}$. The holographic principle seems to imply that the gravitational path integral, however that may be defined, depends only on the data of the asymptotic boundary of the bulk manifold, such as the induced conformal class of metrics and the induced spin structure on $X$.

The independence of the right-hand-side of \eqref{eq:holographic-dictionary} has two immediate consequences. The first and most obvious is that if $\mathcal{M}_1$ and $\mathcal{M}_2$ are bulk manifolds which both have boundary $X$, then
\begin{equation}\label{eq:background-independence-intro}
Z_{\text{grav}}(\mathcal{M}_1)=Z_{\text{grav}}(\mathcal{M}_2)\,.
\end{equation}
This property should be thought of as a version of the \textit{background independence} of quantum gravity. Put simply, the gravitational path integral on a background $\mathcal{M}$ with boundary $X$ should somehow include a sum over all other spacetimes with the same boundary. This sum over geometries is often put into the gravitational path integral by hand. An immediate corollary of \eqref{eq:background-independence-intro} is that if the boundary $X$ is disconnected, then the quantum gravity partition function should factorize, despite the existence of connected on-shell spacetimes with disconnected boundaries \cite{Maldacena:2004rf}.

The second property pertains to spacetimes $\mathcal{M}$ whose boundaries are empty. Such spacetimes are sometimes called `baby universes' in the quantum gravity literature. The AdS/CFT correspondence predicts that the gravitational partition function of such spacetimes must satisfy
\begin{equation}\label{eq:one-state-intro}
Z_{\text{grav}}(\mathcal{M})=1\,.
\end{equation}
This follows immediately from the fact that a CFT on the empty set can only have one state, namely the ground state. Equation \eqref{eq:one-state-intro} implies that the Hilbert space of quantum gravity in a baby universe has precisely one state. The idea that the Hilbert space of quantum gravity in a closed universe is one-dimensional dates back to Coleman \cite{Coleman:1988cy} and has recently gained traction in the context of solvable models of gravity \cite{Marolf:2020xie} and in the swampland program \cite{McNamara:2020uza}.

Understanding the properties \eqref{eq:background-independence-intro} and \eqref{eq:one-state-intro} in the context of concrete UV-complete theories of gravity is a difficult problem which in large part due to the ambiguity in the definition of the quantum gravity path integral. For example, in the case of 2D dilaton gravity, the naive definition of the gravitational path integral does not lead to factorizing partition functions. However, it has been shown that there is a consistent modification of the path integral which does lead to factorizing answers by including nonperturbative extended objects in the sum over topologies \cite{Blommaert:2021fob}. A similar construction leads to a trivial Hilbert space in closed universes \cite{Usatyuk:2024mzs,Usatyuk:2024isz}.

Another example of a UV-complete gravitational theory which is under analytic control is the worldsheet theory dual to the symmetric orbifold $\text{Sym}^N(\mathbb{T}^4)$. The bulk theory is described by type IIB string theory on the background $\mathcal{M}\times\text{S}^3\times\mathbb{T}^4$, supported by $k=1$ units of pure NS-NS flux. Here, $\mathcal{M}$ is any locally-$\text{AdS}_3$ spacetime, i.e. a hyperbolic three-manifold. The duality between this `minimal tension' string theory and the symmetric orbifold $\text{Sym}^N(\mathbb{T}^4)$ has been essentially proven in the framework of perturbative string theory \cite{Eberhardt:2018ouy,Eberhardt:2019ywk,Eberhardt:2020akk,Hikida:2020kil,Eberhardt:2020bgq,Dei:2020zui,Knighton:2020kuh,Eberhardt:2021jvj,Dei:2023ivl}.

Although the worldsheet theory is a sigma model on $\mathcal{M}$, and is thus naively sensitive to the precise structure of the bulk, it was argued in \cite{Eberhardt:2021jvj} that the worldsheet path integral is indeed background independent. The proof of this statement used the classical fact that any hyperbolic three-manifold is a quotient $\mathbb{H}^3/\Gamma$ of hyperbolic three-space by a discrete subgroup $\Gamma\subset\text{SL}(2,\mathbb{C})$. This allows one to treat the worldsheet theory as an orbifold of the theory on $\mathbb{H}^3\times\text{S}^3\times\mathbb{T}^4$, which is well-understood. The proof of background independence then boils down to showing that the worldsheet path integral gives the same answer for two subgroups $\Gamma_1,\Gamma_2$ so long as
\begin{equation}
\partial(\mathbb{H}^3/\Gamma_1)\cong\partial(\mathbb{H}^3/\Gamma_2)\,,
\end{equation}
where $\cong$ denotes conformal equivalence.

In this note, we revisit the arguments of \cite{Eberhardt:2021jvj} using the recently-discovered free field realization of $\mathfrak{psu}(1,1|2)$ explored in \cite{Beem:2023dub,Dei:2023ivl}. We review the worldsheet CFT on $\text{AdS}_3\times\text{S}^3\times\mathbb{T}^4$, paying specific attention to the free field realization of \cite{Beem:2023dub,Dei:2023ivl}, and explain how to define the path integral on $\mathcal{M}\times\text{S}^3\times\mathbb{T}^4$ for any hyperbolic 3-manifold $\mathcal{M}$. We find that the resulting path integral depends only on the conformal data of the asymptotic boundary $X$ of $\mathcal{M}$, and is completely insensitive to the details of the bulk. We also show that the worldsheet path integral localizes to holomorphic maps $\gamma:\Sigma\to X$, in a fashion analogous to the topological A-model.

Since the path integral depends only on the conformal data of the boundary $X$, we conclude immediately that the worldsheet theory is background independent, verifying the result of \cite{Eberhardt:2021jvj} in this new language. We also comment briefly on the case of spacetimes with empty boundary, and argue that the path integral is `empty'. This implies that the string theory free energy on such spacetimes vanishes, and so the gravitational partition function is
\begin{equation}
Z_{\text{grav}}(\mathcal{M})=\exp\left(\mathcal{F}_{\text{st}}(\mathcal{M})\right)=1\,,
\end{equation}
as expected from the holographic principle. Finally, we explain that these properties (background independence and triviality in closed universes) continue to hold true as one deforms continuously away from the tensionless limit by turning on infinitesimal Ramond-Ramond flux.

\section{The minimal tension string}\label{sec:worldsheet-theory}

Type IIB superstrings on $\text{AdS}_3\times\text{S}^3\times\mathbb{T}^4$ can be treated in the RNS formalism, for which the worldsheet CFT is described by an $\mathcal{N}=1$ supersymmetric WZW model of the form
\begin{equation}
\mathfrak{sl}(2,\mathbb{R})_k^{(1)}\oplus\mathfrak{su}(2)_k^{(1)}\oplus\mathbb{T}^4\,,
\end{equation}
and well as the usual superconformal ghosts. While enabling a straightforward covariant quantization scheme, the RNS description has the disadvantage of obscuring the target space (and thus dual CFT) supersymmetries, and is thus not well-suited for analyzing the bulk/boundary dictionary in AdS/CFT.

An alternative quantization scheme involves the hybrid formalism of Berkovits-Vafa-Witten \cite{Berkovits:1999im}. This formulation trades the $\mathcal{N}=1$ WZW models with a WZW model on the supergroup $\text{PSU}(1,1|2)$, which is the group of left super-isometries of $\text{AdS}_3\times\text{S}^3$. As payment for the gift of manifest target-space supersymmetry, the compact $\mathbb{T}^4$ becomes topologically twisted, and the superconformal ghost system gets traded for the usual $b,c$ ghosts, as well as an exotic new $c=28$ scalar, known as the $\rho$ `ghost'. All told, the worldsheet theory on $\text{AdS}_3\times\text{S}^3\times\mathbb{T}^4$ takes the schematic form
\begin{equation}
\mathfrak{psu}(1,1|2)_k\oplus(\text{topologically twisted }\mathbb{T}^4)\oplus(b,c,\rho)\,.
\end{equation}
Or, in terms of actions,
\begin{equation}
S=S_{\mathfrak{psu}}+S_{\mathbb{T}^4}+S_{b,c}+S_{\rho}\,,
\end{equation}
where $S_{\mathfrak{psu}}$ is the usual WZW model action on the supergroup $\text{PSU}(1,1|2)$. The $\mathbb{T}^4$ action is identical to the usual one, except the fermions are now topologically twisted (i.e. half have conformal weight $h=1$, while the other half have $h=0$). The topologically twisted $\mathbb{T}^4$ has central charge $c=0$, and so the total central charge of this theory is
\begin{equation}
c(\mathfrak{psu}(1,1|2)_k)+c(b,c)+c(\rho)=-2-26+28=0\,,
\end{equation}
and thus describes a critical string theory. Here, we used the fact that the $\text{PSU}(1,1|2)$ WZW model has central charge $c=-2$, independent of the level.

When the amount $k$ of NS-NS flux is taken to its minimal value ($k=1$), the $\text{PSU}(1,1|2)$ WZW model admits a description in terms of a free field theory. There are several equivalent descriptions of this free field theory, but the one which is of most direct use is that of \cite{Beem:2023dub,Dei:2023ivl}. There, it was argued that the $\text{PSU}(1,1|2)$ WZW model is equivalent to a free CFT consisting of
\begin{itemize}

	\item A bosonic first-order system $(\beta,\gamma)$ with $h(\beta)=1$ and $h(\gamma)=0$.

	\item Two fermionic first-order systems $(p_a,\theta^a)$ with $h(p_a)=1$ and $h(\theta^a)=0$ ($a=1,2$).

\end{itemize}
Each of these fields also has a right-moving counterpart. This system has action
\begin{equation}\label{eq:free-field-action}
S_{\mathfrak{psu}}=\frac{1}{2\pi}\int_{\Sigma}\left(\beta\overline{\partial}\gamma+p_a\overline{\partial}\theta^a+\bar{\beta}\partial\bar{\gamma}+\bar{p}_a\partial\bar{\theta}^a\right)\,,
\end{equation}
and has central charge $c=2-4=-2$, which is the central charge of the $\text{PSU}(1,1|2)$ model.

To get an intuition for the target space interpretation of the free fields, we note that the global $\text{PSU}(1,1|2)$ isometry group of the superspace $\text{Super}(\text{AdS}_3\times\text{S}^3)$ is a classical symmetry of the action $S_{\mathfrak{psu}}$. The action on the free fields is described as follows. Consider the supergroup element\footnote{We will always work with the \textit{complexified} group $\text{PSU}(1,1|2)_{\mathbb{C}}$. The choice of real form will correspond to a choice of signature in $\text{AdS}_3$.}
\begin{equation}
g=
\begin{pmatrix}
a & b & h_{11} & h_{12}\\
c & d & h_{21} & h_{22}\\
f_{11} & f_{12} & A & B\\
f_{21} & f_{22} & C & D
\end{pmatrix}\in\text{PSU}(1,1|2)\,,
\end{equation}
where $a,b,c,d$ are entries in an $\text{SL}(2,\mathbb{C})$ matrix, $A,B,C,D$ are entries in an $\text{SU}(2,\mathbb{C})$ matrix, and $h_{ij},f_{ij}$ are Grassman variables generating the $8$ supercharges. The action of $g$ on the free fields $(\gamma,\theta^1,\theta^2)$ is by the fractional linear transformations
\begin{equation}
\begin{split}
g\cdot\gamma&=\frac{a\gamma+b+h_{11}\theta^1+h_{12}\theta^2}{c\gamma+d+h_{21}\theta^1+h_{22}\theta^2}\,,\\
g\cdot\theta^1&=\frac{f_{11}\gamma+f_{12}+A\theta^1+B\theta^2}{c\gamma+d+h_{21}\theta^1+h_{22}\theta^2}\,\\
g\cdot\theta^2&=\frac{f_{21}\gamma+f_{22}+C\theta^1+D\theta^2}{c\gamma+d+h_{21}\theta^1+h_{22}\theta^2}
\end{split}
\end{equation}
For purely bosonic transformations (i.e. with $h_{ij}=f_{ij}=0$), we see that $\text{PSU}(1,1|2)$ acts on $\gamma$ via M\"obius transformations, while the fermions $\theta^a$ live in the defining representation of $\text{SU}(2)$, while transforming like a field of weight $h=-\frac{1}{2}$ under $\text{SL}(2,\mathbb{C})$.

Geometrically, the above discussion suggests a concrete meaning of the free fields $\gamma,\theta^1,\theta^2$, namely that they represent the holomorphic $\mathcal{N}=2$ superspace coordinates on the boundary of $\text{AdS}_3$. Indeed, the above fractional linear transformations are the expected global supersymmetry transformations of a 2D CFT with $\mathcal{N}=4$ supersymmetry. As $\gamma$ is taken to be a complex scalar, the worldsheet theory describes global $\text{AdS}_3$, which in Euclidean signature is the global hyperbolic space $\mathbb{H}^3$, i.e. the solid Poincar\'e ball. Some care is needed to take into consideration the north pole of the boundary sphere, see for example the respective discussions in \cite{Frenkel:2005ku,Dei:2023ivl}.

Once the $\text{PSU}(1,1|2)$ WZW model is defined, the worldsheet path integral is defined in terms of a twisted $\mathcal{N}=4$ algebra on the worldsheet. The details are somewhat complicated and we will not need the specifics, see \cite{Berkovits:1999im,Berkovits:1994vy,Dei:2020zui,Dei:2023ivl} for a detailed discussion. The upshot is that the worldsheet free energy is then defined to be a loop expansion:
\begin{equation}\label{eq:string-free-energy}
\mathcal{F}_{\text{st}}=\sum_{g=0}^{\infty}g_s^{2g-2}\int_{\mathcal{M}_{g}}\left\langle\prod_{\alpha=1}^{g-1}|(G^-,\mu_{\alpha})|^2\prod_{\alpha=g}^{3g-3}|\tilde{G}^-,\mu_{\alpha}|^2\left(\int_{\Sigma}|G^+|^2\right)^{g-1}\int_{\Sigma}|J|^2\right\rangle\,,
\end{equation}
where $G^{\pm},\tilde{G}^{\pm}$ are the $\mathcal{N}=4$ generators on the worldsheet, and $J$ is the Cartan of the worldsheet $\text{SU}(2)$ $R$-symmetry. The expectation value is defined formally by the path integral over the fields $(\beta,\gamma,p_a,\theta^a)$, as well as the fields of the topologically-twisted $\mathbb{T}^4$ and the $(b,c,\rho)$ ghost system.

The string free energy $\mathcal{F}_{\text{st}}$ represents the connected piece of the full second quantized string theory partition function, which we will denote as $Z_{\text{grav}}$, as it will play the role of a `gravitational' partition function in our theory. The relationship between the two is simply
\begin{equation}
Z_{\text{grav}}=\exp\left(\mathcal{F}_{\text{st}}\right)\,.
\end{equation}

\section{Defining the path integral}\label{sec:path-integral}

The above definition of the minimal tension string on $\text{AdS}_3\times\text{S}^3\times\mathbb{T}^4$ was exactly that -- a definition of the worldsheet theory on global (Euclidean) $\text{AdS}_3$. However, the AdS/CFT dictionary relates string theories on asymptotically-AdS spacetimes with arbitrary boundary topologies to CFTs living on the boundary. Thus, in order to say things about more interesting CFTs (i.e. those not defined on the 2-sphere), we need to formulate the worldsheet theory on more interesting bulk manifolds.

Any bulk manifold $\mathcal{M}\times\text{S}^3\times\mathbb{T}^4$ which has a hope of having a CFT dual has to be at least asymptotically hyperbolic, in the sense that $\mathcal{M}$ has an asymptotic boundary near which one can take the metric tensor to have constant negative curvature. If we take $\mathcal{M}$ to be \textit{globally} hyperbolic (in the sense that $\mathcal{M}$ admits a metric of constant negative curvature everywhere), then a classical theorem in the geometry of 3-manifolds guarantees that $\mathcal{M}$ is the quotient of global hyperbolic space $\mathbb{H}^3$ by a discrete group $\Gamma\subset\text{SL}(2,\mathbb{C})$. Thus, one way to describe string theory on a hyperbolic 3-manifold $\mathcal{M}$ is to consider the orbifold of string theory on global $\text{AdS}_3$ by the discrete group $\Gamma$.

This prescription, while in principle correct, is more subtle than it might originally seem. The reason is that while the discrete group $\Gamma$ acts on $\mathbb{H}^3$ nicely, it doesn't necessarily act nicely on the sphere $\text{S}^2\cong\partial\mathbb{H}^3$. In most cases, there will be a nonempty subset $\Omega$ of $\text{S}^2$ for which $\Gamma$ acts `badly', while the complement $U=\text{S}^2\setminus\Omega$ admits a free action under $\Gamma$. The boundary of $\mathbb{H}^3/\Gamma$, then, is not $\text{S}^2/\Gamma$ (which is in general a very singular object), but rather $U/\Gamma$. The set $\Omega$ is guaranteed to be a closed measure-zero set, and is usually extremely complicated and fractal.

Let us nevertheless proceed. In order to define the path integral on $\mathbb{H}^3/\Gamma$, we need to know how a discrete subgroup of $\text{SL}(2,\mathbb{C})$ acts on the worldsheet fields. The generators of the global $\mathfrak{sl}(2,\mathbb{C})$ symmetry on the worldsheet are the charges \cite{Dei:2023ivl}
\begin{equation}
J^+_0=\oint\beta\,,\quad J^3_0=\oint\left((\beta\gamma)+\frac{1}{2}(p_a\theta^a)\right)\,,\quad J^-_0=\oint\left((\beta\gamma^2)+\gamma(p_a\theta^a)\right)\,,
\end{equation}
where $(\cdots)$ denotes normal ordering. By acting on the worldsheet fields with these charges, one can read off the finite transformation laws
\begin{equation}\label{eq:sl2-transformations}
\begin{gathered}
\gamma\to\frac{a\gamma+b}{c\gamma+d}\,,\quad\beta\to(c\gamma+d)^{2}\beta-c(c\gamma+d)(p_a\theta^a)\,,\\
p_a\to(c\gamma+d)p_a\,,\quad\theta^a\to(c\gamma+d)^{-1}\theta^a
\end{gathered}
\end{equation}
for some matrix
\begin{equation}
\begin{pmatrix}
a & b\\ c & d
\end{pmatrix}\in\text{SL}(2,\mathbb{C})\,.
\end{equation}
Defining the worldsheet orbifold $\mathbb{H}^3/\Gamma$ is now a matter of choosing a discrete subgroup $\Gamma\subset\text{SL}(2,\mathbb{C})$ and treating the transformation laws \eqref{eq:sl2-transformations} as equivalence relations in the path integral. As discussed above, this process is poorly defined for $\gamma$, since there will be a measure zero subset $\Omega$ of points on $\text{S}^2$ for which $\Gamma$ acts poorly. To remide this, we define $\gamma$ to live in the set $U=\text{S}^2\setminus\Omega$.\footnote{If $\Omega$ is non-empty, then we can WLOG pick $\infty\in\Omega$, so that $\gamma\neq\infty$ in the path integral. If $\Omega$ is empty, we have to add in the point at infinity by hand. This case is discussed at length in Section 4 of \cite{Dei:2023ivl}.}

While in principle this orbifold is well-defined, the non-homogeneous term in the transformation law of $\beta$ in equation \eqref{eq:sl2-transformations} means that there will be non-trivial mixing between $\beta$ and the fermions $p_a,\theta^a$, making the computation of the path integral somewhat complicated. To fix this, follow the trick of \cite{Frenkel:2008vz}. Namely, we define a new field
\begin{equation}
\beta'=\beta+\omega\,(p_a\theta^a)\,,
\end{equation}
where $\omega$ is a scalar on the worldsheet which transforms under $\text{SL}(2,\mathbb{R})$ transformations as
\begin{equation}\label{eq:omega-transform}
\omega\to \omega+c(c\gamma+d)\,.
\end{equation}
This modification in turn cancels the non-homogeneous term in \eqref{eq:sl2-transformations}, so that $\beta'$ transforms as
\begin{equation}
\beta'\to(c\gamma+d)^2\beta'\,.
\end{equation}
The redefinition of $\beta$ also has the effect of changing the free-field action \eqref{eq:free-field-action} to
\begin{equation}
S=\frac{1}{2\pi}\int_{\Sigma}(\beta'\overline{\partial}\gamma+p_a\overline{\nabla}\theta^a+\bar{\beta}'\partial\bar{\gamma}+\bar{p}_a\nabla\bar{\theta}^a)\,,
\end{equation}
where we have defined the covariant derivative
\begin{equation}
\overline{\nabla}=\overline{\partial}-\omega\overline{\partial}\gamma\,.
\end{equation}
This covariant derivative is defined in such a way that $p_a\overline{\nabla}\theta^a$ is invariant under $\text{SL}(2,\mathbb{R})$ transformations. In this way, we should think of $\omega\overline{\partial}\gamma$ as a connection of the $\Gamma$-bundle on the worldsheet of which $\theta^a$ is a section. We can explicitly construct $\omega$ by picking a metric $\rho(\gamma,\bar{\gamma})\mathrm{d}\gamma\,\mathrm{d}\bar{\gamma}$ in the conformal gauge on the boundary of $\mathbb{H}^3$. Given the transformation law
\begin{equation}
\rho(\gamma,\bar{\gamma})\to|c\gamma+d|^4\rho(\gamma,\bar{\gamma})
\end{equation}
of the boundary metric under $\text{SL}(2,\mathbb{R})$ transformations, we see that the combination
\begin{equation}
\omega\overline{\partial}\gamma=\frac{1}{2}\overline{\partial}\log\rho(\gamma,\bar{\gamma})
\end{equation}
transforms in the appropriate way, and thus can be used to define $\omega$. We note that with this identification, the connection $\omega\overline{\partial}\gamma$ is the pullback of the spin connection on the boundary.

From now on, we drop the prime on $\beta'$ and work with the action
\begin{equation}\label{eq:modified-action}
S=\frac{1}{2\pi}\int_{\Sigma}(\beta\overline{\partial}\gamma+p_a\overline{\nabla}\theta^a+\bar{\beta}\partial\bar{\gamma}+\bar{p}_a\nabla\bar{\theta}^a)
\end{equation}
as defining the free-field realization of $\mathfrak{psu}(1,1|2)$. Under $\text{SL}(2,\mathbb{C})$ transformations, the fields now transform homogeneously as
\begin{equation}\label{eq:sl2-transformations-nice}
\begin{gathered}
\gamma\to\frac{a\gamma+b}{c\gamma+d}\,,\quad\beta\to(c\gamma+d)^{2}\beta\,,\\
p_a\to(c\gamma+d)p_a\,,\quad\theta^a\to(c\gamma+d)^{-1}\theta^a\,.
\end{gathered}
\end{equation}
Defining the path integral is now straightforward. We consider a discrete subgroup $\Gamma\subset\text{SL}(2,\mathbb{C})$ and treat the transformation laws \eqref{eq:sl2-transformations-nice} as identifications. Alternatively, we can view the identifications \eqref{eq:sl2-transformations-nice} as defining the spaces to which our worldsheet fields $\gamma,\beta,p_a,\theta^a$ belong. Specifically, the above orbifold identifies $\gamma$ as the holomorphic coordinates of a map from the worldsheet into $X=\partial\mathcal{M}$. On the other hand, $\beta$ transforms as a $(1,0)$-form on $X$. Thus, we can identify $\gamma,\beta$ as living in the spaces
\begin{equation}\label{eq:beta-gamma-spaces}
\gamma\in\text{Map}(\Sigma\to X)\,,\quad\beta\in\Gamma(K_\Sigma\otimes\gamma^*K_X)\,,
\end{equation}
where $K_{\Sigma}$ and $K_X$ are the canonical bundles on the worldsheet and the conformal boundary of $X$, respectively. This is nothing more than the statement that $\beta$ transforms as a $(1,0)$-form under conformal transformation on both $\Sigma$ and $X$. Here, we use the standard notation that $\Gamma(\mathscr{L})$ denotes the space of sections of a line bundle $\mathcal{L}$.

Next, the transformation laws \eqref{eq:sl2-transformations-nice} for the fermions $p_a,\theta^a$ define the data of the spinor bundle on $X$. That is, we can identify
\begin{equation}\label{eq:theta-p-spaces}
\theta^a\in\Gamma(\gamma^*S_X^{-1})\,,\quad p_a\in\Gamma(K_{\Sigma}\otimes\gamma^*S_X)\,,
\end{equation}
where $S_X$ is the holomorphic spin bundle on $X$. The right-moving fields live in the right-moving analogues of these spaces.

As further evidence for this proposal, we can check the transformation laws of the worldsheet fields under \textit{local} conformal transformations on the boundary $X$. These transformations take the form
\begin{equation}
\begin{split}
\gamma\to\gamma-\varepsilon\gamma^{n+1}\,,&\quad \beta\to(1+\varepsilon(n+1)\gamma^{n})\beta\,,\\
\theta^a\to\left(1-\varepsilon\frac{n+1}{2}\gamma^n\right)\theta^a\,,&\quad p_a\to\left(1-\varepsilon\frac{n+1}{2}\gamma^n\right)p_a\,.
\end{split}
\end{equation}
These transformations leave the action \eqref{eq:modified-action} invariant and are generated by an infinite set of conserved charges $\mathcal{L}_n$ generated by the asymptotic symmetries of $\text{AdS}_3$, and which are dual to the Virasoro algebra in the boundary CFT \cite{Brown:1986nw}. These transformation laws are the infinitesimal version of the finite local conformal transformations
\begin{equation}
\begin{split}
\gamma\to f(\gamma)\,,&\quad\beta\to(f'(\gamma))^{-1}\beta\,,\\
\theta^a\to\sqrt{f'(\gamma)}\,\theta^a\,,&\quad p_a\to\sqrt{f'(\gamma)^{-1}}\,p_a\,.
\end{split}
\end{equation}
These transformation laws correspond precisely to the transformation laws of sections of the bundles \eqref{eq:beta-gamma-spaces} and \eqref{eq:theta-p-spaces}. The ambiguity of defining the square roots in the above transformation laws is captured in the choice of spin structure on the boundary $X$, which we take as input.\footnote{This ambiguity can be traced back to the fact that it is actually $\text{PSL}(2,\mathbb{C})=\text{SL}(2,\mathbb{C})/\{\textbf{1}\sim-\textbf{1}\}$ which acts faithfully on $\mathbb{H}^3$. The ambiguity in the choice of sign in the transformation law for $p_a,\theta^a$ in \eqref{eq:sl2-transformations} is reflected in the choice of spin structure on $X$.}

The result of this discussion is that we can bypass the usual orbifolding procedure of defining the worldsheet theory on $\mathcal{M}=\mathbb{H}^3/\Gamma$ by simply identifying the free fields as belonging to the spaces \eqref{eq:beta-gamma-spaces} and \eqref{eq:theta-p-spaces} in the path integral. Note that this construction is formally very similar to the definition of topological sigma models in two dimensions \cite{Witten:1988xj}, except that there are twice as many fermions, and the fermions transform as spinors (not one-forms) on the target space $X$.

In order to define the path integral measure, we note that the spaces \eqref{eq:beta-gamma-spaces} and \eqref{eq:theta-p-spaces} to which $\gamma,\beta,\theta^a,p_a$ all belong are well-defined and admit canonical metrics. Specifically, if we choose a metric $h$ on $\Sigma$ and $g$ on $X$, we can define the path integral measure implicitly by specifying norms of variations $\delta\gamma,\delta\beta,\delta p_a,\delta\theta^a$. These norms are given by
\begin{equation}
\begin{split}
||\delta\gamma||^2=\int_{\Sigma}\mathrm{d}^2z\sqrt{h}\,g_{x\bar{x}}\delta\gamma\delta\bar{\gamma}\,,&\quad ||\delta\beta||^2=\int_{\Sigma}\mathrm{d}^2z\sqrt{h}\,h^{z\bar{z}}g^{x\bar{x}}\delta\beta\,\delta\bar{\beta}\,,\\
||\delta\theta^a||^2=\int_{\Sigma}\mathrm{d}^2z\sqrt{h}\sqrt{g_{x\bar{x}}}\,\delta\theta^a\delta\bar{\theta}^a\,,&\quad ||\delta p_a||^2=\int_{\Sigma}\mathrm{d}^2z\sqrt{h}\,h^{z\bar{z}}\sqrt{g^{x\bar{x}}}\,\delta p_a\delta\bar{p}_a\,.
\end{split}
\end{equation}
Note that the definitions of these norms are not Weyl-invariant on the worldsheet, and so the path integral measure will have a Weyl anomaly. This will be canceled by the central charge of the $c=2$ ghost system, so that the full path integral is Weyl-invariant.

\section{The path integral localizes}\label{sec:localization}

A key feature of the minimal tension string is that its path integral localizes to a discrete sum of holomorphic covering spaces of $X$. This was shown in \cite{Eberhardt:2019ywk,Dei:2020zui,Knighton:2020kuh} for $X\cong\text{S}^2$ and extended in \cite{Eberhardt:2020bgq,Eberhardt:2021jvj} to boundaries of higher genus. These proofs used a somewhat complicated analysis based on a different free-field realization than the one presented here. In the free field representation with action \eqref{eq:modified-action}, this argument can be simplified by a great deal. Let us briefly explain how this works.

Assuming that none of the vertex operators in a correlation function depend on $\beta$, we can integrate it out. The result is that the path integral includes a delta-functional $\delta^{(2)}(\overline{\partial}\gamma)$ which imposes that the map $\gamma:\Sigma\to X$ is holomorphic. The space $\mathcal{H}(\Sigma\to X)$ of holomorphic maps is disconnected, and breaks up into a union of connected spaces which depend on the degree $\text{deg}(\gamma)$ of the map (i.e. the number of preimages of a generic point). This component of $\mathcal{H}(\Sigma\to X)$ has dimension
\begin{equation}
\text{deg}(\gamma)(2-2G)+1-g\,,
\end{equation}
so that the dimension of the full path integral (including the integration over $\mathcal{M}_g$) is
\begin{equation}\label{eq:path-integral-dimension}
\text{deg}(\gamma)(2-2G)+2g-2\,.
\end{equation}
Now, the degree of the map $\gamma$ in the path integral is not allowed to take any random value, but is constrained by the zero mode condition of the fermions $p_a,\theta^a$. The supercharges $\tilde{G}^-$ appearing in the definition of the free energy \eqref{eq:string-free-energy} are proportional to $p_1p_2$ \cite{Dei:2023ivl} and so the number of, say, $p_1$'s in the correlator defining $\mathcal{F}_{\text{st}}$ is $2g-2$. However, the number of zero modes of $\theta_1$ minus the number of zero modes of $p^1$ is given by the index of the covariant derivative $\overline{\nabla}$, which is readily calculated by the Riemann-Roch theorem:
\begin{equation}
\begin{split}
\text{dim}\,\text{ker}(\overline\nabla)-\text{dim}\,\text{coker}(\overline{\nabla})&=\deg(\gamma^*(S_X))+1-g\\
&=\text{deg}(\gamma)(1-G)+1-g\,.
\end{split}
\end{equation}
Demanding that this equal $2g-2$, so that the path integral \eqref{eq:string-free-energy} is nonzero, gives the selection rule
\begin{equation}\label{eq:selection-rule}
\text{deg}(\gamma)(1-G)=1-g
\end{equation}
which completely determines the degree of the map $\gamma$ in terms of the worldsheet and boundary genera.

By the Riemann-Hurwitz formula, the selection rule \eqref{eq:selection-rule} implies that the path integral can only include holomorphic maps $\gamma$ which are topologically covering maps (that is, $\gamma$ has no branch points). Plugging the selection rule \eqref{eq:selection-rule} into the dimension \eqref{eq:path-integral-dimension} of the path integral immediately yields the result that the worldsheet path integral, after integrating out $\beta$ and the fermions $p_a,\theta^a$, localizes to a zero-dimensional integral, i.e. a discrete sum. This sum is over the space of (connected) unbranched holomorphic maps $\gamma$.

It is worth pointing out that the localization principle described above is essentially the same as the analogous property of the topological A-model \cite{Witten:1988xj,Witten:1991zz}, and indeed, as mentioned in the previous section, the path integral of the $k=1$ theory is structurally similar to the A-model at the level of the free fields \eqref{eq:modified-action}. It has been suggested in the literature that the minimal tension string should, in some sense, be related to a topological string theory \cite{Eberhardt:2018ouy,Dei:2020zui}. The above analysis sheds light on the sense in which this is true.

\section{Background independence}\label{sec:background}

In the previous section we argued that the minimal tension string theory on $\mathcal{M}\times\text{S}^3\times\mathbb{T}^4$ is given by a worldsheet CFT whose action is simply the free field action in \eqref{eq:modified-action} (plus the action of the $\mathbb{T}^4$ and ghosts), but for which the fields are taken to lie in sections of various line bundles. Importantly, the only ingredients going into defining this worldsheet CFT are
\begin{itemize}

	\item The complex structure on $\partial\mathcal{M}$, which is induced by the asymptotic form of the metric on $\mathcal{M}$ near the boundary.

	\item The spin structure on $\partial\mathcal{M}$, which simply tells us the periodicity of the fermions $\theta^a$ as we traverse a noncontractible cycle on the boundary.

\end{itemize}
The key observation is that this information is completely independent of the details of the bulk manifold $\mathcal{M}$ which are not visible near the asymptotic boundary. This includes, for example, whether a non-contractible cycle on $\partial\mathcal{M}$ becomes contractible in the bulk.

The structure of the worldsheet CFT seems to imply then that string theory on $\mathcal{M}\times\text{S}^3\times\mathbb{T}^4$ is completely independent of the bulk details of the 3-manifold $\mathcal{M}$, as the fundamental fields are completely specified by asymptotic data on the boundary. In particular, this implies that if $\mathcal{M}$ and $\mathcal{M}'$ are two different bulk manifolds with the same asymptotic boundary data, then any string theory observable computed on $\mathcal{M}\times\text{S}^3\times\mathbb{T}^4$ is equal to the same observable computed on $\mathcal{M}'\times\text{S}^3\times\mathbb{T}^4$.\footnote{We emphasize that we had to use the special property of string theory at $k=1$ units of NS-NS flux that it is described by the free-field action \eqref{eq:free-field-action}. For other values of the string tension, this argument would not go through.} This statement is a concrete realization of the idea of background independence of string theory.

An immediate consequence of this statement is that if $\mathcal{M}$ is a connected 3-manifold with a disconnected boundary $X_1\sqcup\cdots\sqcup X_{\ell}$, then the string free energy \eqref{eq:string-free-energy} satisfies
\begin{equation}
\mathcal{F}_{\text{st}}(\mathcal{M})=\sum_{i=1}^{\ell}\mathcal{F}_{\text{st}}(\mathcal{M}_{i})\,,
\end{equation}
where $\mathcal{M}_i$ are $\ell$ 3-manifolds with $\partial\mathcal{M}_i=X_i$. This can be seen directly from the string path integral: the path integral over $\mathcal{M}$ decomposes decomposes into a sum over free field theories of the form \eqref{eq:free-field-action} with
\begin{equation}
\begin{split}
(\gamma,\bar{\gamma})\in\text{Map}(\Sigma\to X_i)\,,&\quad\beta\in \Gamma(K_{\Sigma}\otimes\gamma^*(K_X))\,,\\
\theta^a\in\Gamma(\gamma^*(S_X^{-1}))\,,&\quad p_a\in \Gamma(K_{\Sigma}\otimes\gamma^*(S_X))\,.
\end{split}
\end{equation}
Exponentiating $\mathcal{F}_{\text{st}}(\mathcal{M})$ to get the gravitational partition function gives
\begin{equation}\label{eq:factorization}
Z_{\text{grav}}(\mathcal{M})=\prod_{i=1}^{\ell}Z_{\text{grav}}(\mathcal{M}_{i})\,.
\end{equation}
That is, the quantum gravity partition function in the $k=1$ string theory \textit{factorizes} when the bulk manifold has a disconnected boundary.

Another consequence is that if $\mathcal{M}$ is a 3-manifold with empty boundary, then the string partition function precisely vanishes. Indeed, if $\mathcal{N}$ is a 3-manifold with boundary $X$, then the disjoint union $\mathcal{N}\sqcup\mathcal{M}$ also has boundary $X$. We thus conclude that
\begin{equation}
\mathcal{F}_{\text{st}}(\mathcal{N}\sqcup\mathcal{M})=\mathcal{F}_{\text{st}}(\mathcal{N})\,,
\end{equation}
so long as $\mathcal{M}$ is compact. However, since $\mathcal{F}_{\text{st}}(\mathcal{N}\sqcup\mathcal{M})=\mathcal{F}_{\text{st}}(\mathcal{N})+\mathcal{F}_{\text{st}}(\mathcal{M})$ (which must be true of \textit{any} string theory), we conclude that\footnote{This can also be seen as an immediate consequence of the fact that $\gamma\in\text{Map}(\Sigma\to\emptyset)=\emptyset$, so that the path integral is empty.}
\begin{equation}
\mathcal{F}_{\text{st}}(\mathcal{M})=0\text{ if }\partial\mathcal{M}=\emptyset\,.
\end{equation}
In terms of the gravitational partition function, which includes disconnected worldsheets, we conclude
\begin{equation}\label{eq:one-state}
Z_{\text{grav}}(\mathcal{M})=1\text{ if }\partial\mathcal{M}=\emptyset\,.
\end{equation}
The implication, then, is that the bulk theory of quantum gravity described by $k=1$ string theory on $\text{AdS}_3\times\text{S}^3\times\mathbb{T}^4$ has precisely one state -- the ground state -- when the bulk manifold is closed.

\section{Beyond minimal tension}\label{sec:deformation}

The discussion of the previous sections relied heavily on the fact that the string theory is described by the $k=1$ limit of the $\text{PSU}(1,1|2)$ WZW model, which is a worldsheet CFT that only depends on the boundary data of the target space. However, one may argue that this is a miracle of minimal tension, and this feature of the string theory may be `lifted' as one deforms away from this critical value of the string tension. Here, we will argue that the background independence is at least preserved for a set of continuous perturbations of the worldsheet theory corresponding to turning on Ramond-Ramond flux.

From the point of view of the dual CFT, the appropriate deformation is by the supersymmetric descendant of a twist-2 field in the symmetric orbifold theory. Specifically, if $\Psi^{\alpha\dot\alpha}_2$ is the BPS field in the $w=2$ twisted sector of the symmetric orbifold, where $\alpha,\dot\alpha$ are spinor indices of the $\mathfrak{su}(2)_L\oplus\mathfrak{su}(2)_R$ R-symmetry, then there is a four-dimensional family of exactly marginal deformations of $\text{Sym}^N(\mathbb{T}^4)$ given by
\begin{equation}\label{eq:spacetime-deformations}
\Phi^{A\dot A}_2(x,\bar{x})=\varepsilon_{\alpha\beta}\varepsilon_{\dot\alpha\dot\beta}\mathcal{G}^{\beta A}_{-\frac{1}{2}}\mathcal{G}^{\dot\beta\dot A}_{-\frac{1}{2}}\Psi^{\alpha\dot\alpha}_2(x,\bar{x})\,.
\end{equation}
Here, $\mathcal{G}^{\alpha A},\overline{\mathcal{G}}^{\dot\alpha\dot A}$ are the $\mathcal{N}=(4,4)$ supersymmetry generators, and the capital roman indices label the representation with respect to the outer $\mathfrak{su}(2)_L\oplus\mathfrak{su}(2)_R$ symmetry of the small $\mathcal{N}=(4,4)$ supersymmetry algebra. These marginal operators, along with the $4\times 4=16$ Narain moduli of the $\mathbb{T}^4$ lattice, parametrize the 20 dimensional moduli space which contains $\text{Sym}^N(\mathbb{T}^4)$.

On the worldsheet, the operators \eqref{eq:spacetime-deformations} are dual to exactly marginal operators in the worldsheet sigma model which take the form
\begin{equation}\label{eq:worldsheet-deformations}
\mathcal{O}^{A\dot A}_2(z,\bar{z})=\tilde{G}\,^{-}_{-1}\tilde{\overline{G}}\,^{-}_{-1}\int_X\varepsilon_{\alpha\beta}\varepsilon_{\dot\alpha\dot\beta}\mathcal{G}^{\beta A}_{-\frac{1}{2}}\mathcal{G}^{\dot\beta\dot A}_{-\frac{1}{2}}\mathcal{V}_{2}^{\alpha\dot\alpha}(z,\bar{z},x,\bar{x})\,,
\end{equation}
where $\mathcal{V}_2^{\alpha\dot\alpha}$ is the worldsheet dual of $\Psi_2^{\alpha\dot\alpha}$ and the operators $\mathcal{G}$ should be thought of as the corresponding DDF operators on the worldsheet which were constructed in \cite{Naderi:2022bus}. It was shown in \cite{Fiset:2022erp} that the worldsheet deformation by $\mathcal{O}_2^{A\dot A}$ is equivalent to deforming by $\Phi_2^{A\dot A}$ in the boundary CFT to second order in conformal perturbation theory.\footnote{This class of deformations have also been recently shown to admit an integrable structure in the boundary CFT, see \cite{Gaberdiel:2023lco}.}

Let us briefly argue that how this result structurally holds at all orders in perturbation theory. Firstly, the BPS operators $\mathcal{V}_2^{\alpha\dot\alpha}$ can be written in closed form as \cite{Dei:2023ivl}
\begin{equation}
\mathcal{V}_2^{\alpha\dot\alpha}=e^{2\rho+i\sigma+iH}\partial\theta^{\alpha}\overline{\partial}\bar{\theta}^{\dot\alpha}\overline{\theta}^1\overline{\theta}^2\theta^1\theta^2\delta^{(2)}_2(\gamma(z),x)\,,
\end{equation}
where $\alpha,\dot\alpha\in\{1,2\}$. Here, we have defined the delta function
\begin{equation}
\delta^{(2)}_2(\gamma(z),x):=\delta(\gamma(z),x)\delta(\partial\gamma(z))\delta(\bar{\gamma}(\bar{z}),\bar{x})\delta(\overline{\partial}\bar{\gamma}(\bar{z}))\,,
\end{equation}
which is a $(-1,-1)$-form valued distribution in the path integral that demands the criticality condition
\begin{equation}
\gamma(y)\sim x+\mathcal{O}((y-z)^2)\,,\quad y\to z\,.
\end{equation}
Now, let us ignore the various supercharges appearing in equation \eqref{eq:worldsheet-deformations} as well as the various indices. At $m^{\text{th}}$ order in worldsheet perturbation theory, we consider the correlator
\begin{equation}
\int_{\mathcal{M}_{g,n}}\Braket{\left(\int\mathcal{O}_2\right)^m\cdots}\,,
\end{equation}
where $\cdots$ represents the fields involved in the undeformed correlation function. The effect of inserting $\mathcal{O}_2$ in the correlation function is to introduce $m$ critical points $y_i$ on the worldsheet such that
\begin{equation}
\partial\gamma(z)\sim\mathcal{O}(z-y_i)\,,\quad z\to y_i\,,
\end{equation}
where the image $\gamma(y_i)$ is left arbitrary, i.e. is integrated over in the path integral. Put in other words, if the undeformed correlation function is a path integral over holomorphic maps $\gamma$ with prescribed behavior, the deformed path integral at $m^{\text{th}}$ order in perturbation theory will be an integral over holomorphic maps $\gamma$ with $m$ `extra' simple branch points. This is precisely the prescription for computing perturbation theory with respect to twist-2 fields in the symmetric orbifold \cite{Lunin:2000yv}.

Crucially, every ingredient in the worldsheet path integral is a function only of worldsheet data. The worldsheet vertex operator dual to the spacetime BPS state lives in the space
\begin{equation}
\mathcal{V}_2^{\alpha\dot\alpha}\in\Gamma(\gamma^*(S_X\otimes\overline{S}_X))\,.
\end{equation}
The deformation operator is built out of $\mathcal{V}_2^{\alpha\dot\alpha}$ and the DDF operators
\begin{equation}
\mathcal{G}_{-\frac{1}{2}}^{\alpha A}\in\Gamma(\gamma^*(S_X))\,,\quad \mathcal{G}_{-\frac{1}{2}}^{\alpha A}\in\Gamma(\gamma^*(\overline{S}_X))\,,
\end{equation}
as well as an integral over $X$. Since none of these ingredients depend in any way on the details of the three-dimensional bulk, we conclude that background independence is preserved even as we deform away from the orbifold point. Of course, this is to be expected if one believes that the AdS/CFT correspondence holds in the full moduli space containing $\text{Sym}^N(\mathbb{T}^4)$.

\section{Discussion}

In this note, we argued that the gravitational partition function of minimal tension string on $\mathcal{M}\times\text{S}^3\times\mathbb{T}^4$ depends only on asymptotic information encoded on the boundary of $\mathcal{M}$ such as the induced conformal class of metrics and the induced spin structure. This implies that the second-quantized string theory is background independent. As a corollary, we argued that the quantum gravity Hilbert space on a closed hyperbolic three-manifold $\mathcal{M}$ contains exactly one state.

The arguments in this paper, however, were extremely specific to the case of $k=1$ units of NS-NS flux, where the worldsheet theory admits an extremely powerful free field realization. While we argued that the background independence should remain upon continuous deformations of the worldsheet theory (e.g. when switching on infinitesimal RR flux), it is unclear how the arguments in this paper would extend to $k>1$. For larger values of $k$, the definition of the worldsheet sigma model depends intimately on the details of the bulk manifold away from the boundary. Thus, while we expect background independence to remain true for larger values of NS-NS flux, the mechanism whereby background independence appears will likely be much more complicated and will most likely be invisible from string perturbation theory alone.

\subsection*{Acknowledgements} 
I thank Andrea Dei, Matthias Gaberdiel, Kiarash Naderi, Vit Sriprachyakul, Jakub Vo\v{s}mera and especially Mykhaylo Usatyuk for stimulating discussions. I would also like to thank thank Matthias Gaberdiel and Mykhaylo Usatyuk for helpful comments on a draft of this manuscript, as well as the oranizers of the KITP program ``What is string theory? Weaving together perspectives.'' for their hospitality during the writing of this work. This research was supported by STFC consolidated grants ST/T000694/1 and ST/X000664/1, and was supported in part by grant NSF PHY-2309135 to the Kavli Institute for Theoretical Physics (KITP).

\bibliography{references.bib}

\providecommand{\href}[2]{#2}\begingroup\raggedright\begin{thebibliography}{10}

\bibitem{Maldacena:1997re}
J.~M. Maldacena, ``{The Large $N$ Limit of Superconformal Field Theories and
  Supergravity},'' {\em Int. J. Theor. Phys.} {\bf 38} (1999) 1113--1133,
  \href{http://www.arXiv.org/abs/hep-th/9711200}{{\tt hep-th/9711200}}.
[Adv. Theor. Math. Phys.2,231(1998)].

\bibitem{Witten:1998qj}
E.~Witten, ``{Anti-de Sitter space and holography},'' {\em Adv. Theor. Math.
  Phys.} {\bf 2} (1998) 253--291,
  \href{http://www.arXiv.org/abs/hep-th/9802150}{{\tt hep-th/9802150}}.

\bibitem{Maldacena:2004rf}
J.~M. Maldacena and L.~Maoz, ``{Wormholes in AdS},'' {\em JHEP} {\bf 02} (2004)
  053, \href{http://www.arXiv.org/abs/hep-th/0401024}{{\tt hep-th/0401024}}.

\bibitem{Coleman:1988cy}
S.~R. Coleman, ``{Black holes as red herrings: Topological fluctuations and the
  loss of quantum coherence},'' {\em Nucl. Phys. B} {\bf 307} (1988) 867--882.

\bibitem{Marolf:2020xie}
D.~Marolf and H.~Maxfield, ``{Transcending the ensemble: baby universes,
  spacetime wormholes, and the order and disorder of black hole information},''
  {\em JHEP} {\bf 08} (2020) 044,
  \href{http://www.arXiv.org/abs/2002.08950}{{\tt 2002.08950}}.

\bibitem{McNamara:2020uza}
J.~McNamara and C.~Vafa, ``{Baby Universes, Holography, and the Swampland},''
  \href{http://www.arXiv.org/abs/2004.06738}{{\tt 2004.06738}}.

\bibitem{Blommaert:2021fob}
A.~Blommaert, L.~V. Iliesiu, and J.~Kruthoff, ``{Gravity factorized},'' {\em
  JHEP} {\bf 09} (2022) 080, \href{http://www.arXiv.org/abs/2111.07863}{{\tt
  2111.07863}}.

\bibitem{Usatyuk:2024mzs}
M.~Usatyuk, Z.-Y. Wang, and Y.~Zhao, ``{Closed universes in two dimensional
  gravity},'' \href{http://www.arXiv.org/abs/2402.00098}{{\tt 2402.00098}}.

\bibitem{Usatyuk:2024isz}
M.~Usatyuk and Y.~Zhao, ``{Closed universes, factorization, and ensemble
  averaging},'' \href{http://www.arXiv.org/abs/2403.13047}{{\tt 2403.13047}}.

\bibitem{Eberhardt:2018ouy}
L.~Eberhardt, M.~R. Gaberdiel, and R.~Gopakumar, ``{The Worldsheet Dual of the
  Symmetric Product CFT},'' {\em JHEP} {\bf 04} (2019) 103,
  \href{http://www.arXiv.org/abs/1812.01007}{{\tt 1812.01007}}.

\bibitem{Eberhardt:2019ywk}
L.~Eberhardt, M.~R. Gaberdiel, and R.~Gopakumar, ``{Deriving the
  AdS$_{3}$/CFT$_{2}$ correspondence},'' {\em JHEP} {\bf 02} (2020) 136,
  \href{http://www.arXiv.org/abs/1911.00378}{{\tt 1911.00378}}.

\bibitem{Eberhardt:2020akk}
L.~Eberhardt, ``{$\mathrm{AdS}_{3}/\mathrm{CFT}_{2}$ at higher genus},'' {\em
  JHEP} {\bf 05} (2020) 150, \href{http://www.arXiv.org/abs/2002.11729}{{\tt
  2002.11729}}.

\bibitem{Hikida:2020kil}
Y.~Hikida and T.~Liu, ``{Correlation functions of symmetric orbifold from
  $\mathrm{AdS}_{3}$ string theory},'' {\em JHEP} {\bf 09} (2020) 157,
  \href{http://www.arXiv.org/abs/2005.12511}{{\tt 2005.12511}}.

\bibitem{Eberhardt:2020bgq}
L.~Eberhardt, ``{Partition functions of the tensionless string},'' {\em JHEP}
  {\bf 03} (2021) 176, \href{http://www.arXiv.org/abs/2008.07533}{{\tt
  2008.07533}}.

\bibitem{Dei:2020zui}
A.~Dei, M.~R. Gaberdiel, R.~Gopakumar, and B.~Knighton, ``{Free field
  world-sheet correlators for ${\rm AdS}_3$},'' {\em JHEP} {\bf 02} (2021) 081,
  \href{http://www.arXiv.org/abs/2009.11306}{{\tt 2009.11306}}.

\bibitem{Knighton:2020kuh}
B.~Knighton, ``{Higher genus correlators for tensionless AdS$_{3}$ strings},''
  {\em JHEP} {\bf 04} (2021) 211,
  \href{http://www.arXiv.org/abs/2012.01445}{{\tt 2012.01445}}.

\bibitem{Eberhardt:2021jvj}
L.~Eberhardt, ``{Summing over Geometries in String Theory},'' {\em JHEP} {\bf
  05} (2021) 233, \href{http://www.arXiv.org/abs/2102.12355}{{\tt 2102.12355}}.

\bibitem{Dei:2023ivl}
A.~Dei, B.~Knighton, and K.~Naderi, ``{Solving AdS$_3$ string theory at minimal
  tension: tree-level correlators},''
  \href{http://www.arXiv.org/abs/2312.04622}{{\tt 2312.04622}}.

\bibitem{Beem:2023dub}
C.~Beem and A.~E.~V. Ferrari, ``{Free field realisation of boundary vertex
  algebras for Abelian gauge theories in three dimensions},''
  \href{http://www.arXiv.org/abs/2304.11055}{{\tt 2304.11055}}.

\bibitem{Berkovits:1999im}
N.~Berkovits, C.~Vafa, and E.~Witten, ``{Conformal field theory of AdS
  background with Ramond-Ramond flux},'' {\em JHEP} {\bf 03} (1999) 018,
  \href{http://www.arXiv.org/abs/hep-th/9902098}{{\tt hep-th/9902098}}.

\bibitem{Frenkel:2005ku}
E.~Frenkel and A.~Losev, ``{Mirror symmetry in two steps: A-I-B},'' {\em
  Commun. Math. Phys.} {\bf 269} (2006) 39--86,
  \href{http://www.arXiv.org/abs/hep-th/0505131}{{\tt hep-th/0505131}}.

\bibitem{Berkovits:1994vy}
N.~Berkovits and C.~Vafa, ``{N=4 topological strings},'' {\em Nucl. Phys. B}
  {\bf 433} (1995) 123--180,
  \href{http://www.arXiv.org/abs/hep-th/9407190}{{\tt hep-th/9407190}}.

\bibitem{Frenkel:2008vz}
E.~Frenkel, A.~Losev, and N.~Nekrasov, ``{Instantons beyond topological theory
  II},'' \href{http://www.arXiv.org/abs/0803.3302}{{\tt 0803.3302}}.

\bibitem{Brown:1986nw}
J.~D. Brown and M.~Henneaux, ``{Central Charges in the Canonical Realization of
  Asymptotic Symmetries: An Example from Three-Dimensional Gravity},'' {\em
  Commun. Math. Phys.} {\bf 104} (1986) 207--226.

\bibitem{Witten:1988xj}
E.~Witten, ``{Topological Sigma Models},'' {\em Commun. Math. Phys.} {\bf 118}
  (1988) 411.

\bibitem{Witten:1991zz}
E.~Witten, ``{Mirror manifolds and topological field theory},'' {\em AMS/IP
  Stud. Adv. Math.} {\bf 9} (1998) 121--160,
  \href{http://www.arXiv.org/abs/hep-th/9112056}{{\tt hep-th/9112056}}.

\bibitem{Naderi:2022bus}
K.~Naderi, ``{DDF operators in the hybrid formalism},'' {\em JHEP} {\bf 12}
  (2022) 043, \href{http://www.arXiv.org/abs/2208.01617}{{\tt 2208.01617}}.

\bibitem{Fiset:2022erp}
M.-A. Fiset, M.~R. Gaberdiel, K.~Naderi, and V.~Sriprachyakul, ``{Perturbing
  the symmetric orbifold from the worldsheet},'' {\em JHEP} {\bf 07} (2023)
  093, \href{http://www.arXiv.org/abs/2212.12342}{{\tt 2212.12342}}.

\bibitem{Gaberdiel:2023lco}
M.~R. Gaberdiel, R.~Gopakumar, and B.~Nairz, ``{Beyond the Tensionless Limit:
  Integrability in the Symmetric Orbifold},''
  \href{http://www.arXiv.org/abs/2312.13288}{{\tt 2312.13288}}.

\bibitem{Lunin:2000yv}
O.~Lunin and S.~D. Mathur, ``{Correlation Functions for $ M^N/S^N$
  Orbifolds},'' {\em Commun. Math. Phys.} {\bf 219} (2001) 399--442,
  \href{http://www.arXiv.org/abs/hep-th/0006196}{{\tt hep-th/0006196}}.

\end{thebibliography}\endgroup
\bibliographystyle{utphys.bst}

\end{document}